\documentclass[notitlepage,reqno,a4paper]{article}
\usepackage{mathrsfs}
\usepackage{}
\usepackage{amsfonts}
\usepackage{subfigure}
\usepackage{cite}
\usepackage{stmaryrd}
\usepackage{amsmath}
\usepackage{amssymb}
\usepackage{graphicx}
\usepackage{verbatim}
\usepackage{indentfirst}
\usepackage{txfonts}

\begin{document}
\begin{center}\large \textbf{Scattering problems in the fractional quantum mechanics governed by the 2D space fractional Schr\"odinger equation}
\end{center}
\begin{center}\small {Jianping Dong\footnote{Email:Dongjp.sdu@gmail.com}}\\
{\itshape Department of Mathematics, College of Science, Nanjing
University of Aeronautics and Astronautics, Nanjing 210016,
China}\end{center}
\begin{quote}  The 2D space-fractional Schr\"odinger equation in the time-independent and time-dependent cases for the scattering
problem in the fractional quantum mechanics is studied. We define and give the mathematical expression of the Green's functions for the two cases.
The asymptotic formulas of the Green's functions are also given, and
applied  to get the approximate wave functions for the fractional
quantum scattering problems.\end{quote}
\section{ {Introduction}}
 Nowadays, the fractional calculus \cite{oldham,kilbas} becomes a useful tools for scientists. It has been successfully applied in anomalous transport, diffusion-reaction processes, super-slow relaxation, etc \cite{metzler,west}.  Recently, the fractional calculus enters the world of quantum
 mechanics. In quantum physics, Feynman and Hibbs \cite{feynman}
reformulated the famous Schr\"odinger equation \cite{levin,grif} by
use of the path integral approach considering the Gaussian
probability distribution. The L\'evy stochastic process is a natural
generalization of the Gaussian process. The possibility of
developing the path integral over the paths of the L\'evy motion was
discussed by Kac \cite{kac}, who pointed out that the L\'evy path
integral generates the functional measure in the space of left (or
right) continued functions having only discontinuities of the first
kind. Recently, Laskin\cite{laskin1,laskin3,laskin5} generalized
Feynman path integral to L\'evy one, and developed a
space-fractional Schr\"odinger equation containing the Riesz
fractional derivative \cite{oldham,kilbas}. Then, he constructed the
fractional quantum mechanics and showed some properties of the space
fractional quantum system \cite{laskin2,laskin4}. Afterwards, the
time-fractional, and space-time-fractional Schr\"odinger equation
\cite{naber,wang,dong1} were also given. This paper focuses on the
space fractional quantum systems described by the fractional
Schr\"odinger equation (FSE) given by Laskin \cite{laskin3}. Some
progresses have been made in this field. The solutions to the FSE
with some specific potentials were obtained in
Refs.~\cite{guo,dong2,dong3,Oliveira,laskin4}. The Green's function for the
time-independent 1D FSE of the free particle was also given by Guo
and Xu  in Ref.~\cite{guo}. A generalized space-time-fractional case
was studied by Wang and Xu in Ref.~\cite{wang}.
\par So far, the fractional
Schr\"odinger equation has been studied mainly in the one-dimensional case.
However, in the real world, the higher dimension (2D or 3D) should
be considered, and the exact solution to this equation is often hard
to obtain. In Refs.\cite{dong5,dong6}, the fractional
Schr\"odinger equation with 3D space potential in the
time-independent and time-dependent cases has been studied. A generalized Lippmann-Schwinger equation for the fractional quantum
mechanics was given and applied to get the approximate scattering wave
function of every order for the fractional quantum scattering problem. In quantum mechanics,  the scattering theory
\cite{levin,Shankar} is often used to study the inner structure of a
matter, so the research on the generalized quantum scattering
problem under the framework of fractional quantum mechanics is
meaningful. In the paper, we turn to the two-dimensional case. 2D systems of particles are rapidly becoming of more practical
importance as their realization in surface physics becomes
increasingly easy \cite{robinett}. It is now possible, using modern crystal growth
techniques such as molecular- beam epitaxy and other methods, to
fabricate semiconductor nano-structures, artificially created
patterns of atoms whose atomic composition and sizes are
controllable at the nanometer scale, which is comparable to
interatomic distances. At such length scales, quantum effects
obviously become increasingly important. Even more dramatically,
scanning tunnel microscopy (STM) techniques can now be used to
manipulate individual atoms and molecules with atomic scale
precision. So the 2D fractional quantum mechanics deserves our investigation.
 In this paper, we study the integral form of the fractional
Schr\"odinger equation with 2D space potential, and apply it to
study the fractional quantum scattering problem.

\section{ {Time-dependent Case}} \label{sec1}  The space-fractional Schr\"odinger equation
\cite{laskin3} obtained by Laskin reads (in two dimensions)
\begin{equation} i\hslash\frac{\partial\psi(\mathrm{\textbf{r}},t)}{\partial
t}=H_\alpha \psi(\mathrm{\textbf{r}},t), \label{TDfse1}
\end{equation}
where $\psi(\mathrm{\textbf{r}},t)$ is the time-dependent wave
function , and $H_\alpha \text{ }(1<\alpha\leq 2)$ is the fractional
Hamiltonian operator given by
\begin{equation}
H_\alpha=-D_{\alpha}
(\hslash\nabla)^{\alpha}+V(\mathrm{\textbf{r}},t). \label{TDfse2}
\end{equation}
Here $D_{\alpha}$ with physical dimension
$[D_{\alpha}]=\text{[Energy]}^{1-\alpha}\times
\text{[Length]}^\alpha \times \text{[Time]}^{-\alpha}$ is dependent
on $\alpha$ [$D_{\alpha}=1/2m$ for $\alpha=2$, $m$ denotes the mass
of a particle] and $(\hslash\nabla)^{\alpha}$ is the quantum Riesz
fractional operator \cite{kilbas,laskin1} defined by
\begin{equation}
(\hslash\nabla)^{\alpha}\psi(\mathrm{\textbf{r}},t)=-\frac1{(2\pi\hslash)^2}\int\mathrm{d}^2\mathrm{\textbf{p}}\text{e}^
{i\mathrm{\textbf{p}}\cdot\mathrm{\textbf{r}}/\hslash}|\mathrm{\textbf{p}}|^\alpha\int\text{e}^
{-i\mathrm{\textbf{p}}\cdot\mathrm{\textbf{r}}/\hslash}\psi(\mathrm{\textbf{r}},t)\mathrm{d}^2\mathrm{\textbf{r}}
\label{TDriesz1}.
\end{equation}
Note that by use of the method of dimensional analysis we have given
a specific expression of $D_{\alpha}$ in Ref.~\cite{dong3} as
$D_{\alpha}=\bar{c}^{2-\alpha}/(\alpha m^{\alpha-1})$, where
$\bar{c}$ denotes the characteristic velocity of the
non-relativistic quantum system.

 Now, we define a Green's function of the FSE by
\begin{equation}
[i\hslash\frac{\partial}{\partial t}+D_{\alpha}
(\hslash\nabla)^{\alpha}]G(\mathrm{\textbf{r}},t;\mathrm{\textbf{r}}^\prime,t^\prime)
=\delta(\textbf{r}-\mathrm{\textbf{r}}^\prime)\delta(t-t^\prime),\label{TDgreen}
\end{equation}with the causality condition
\begin{equation}
G(\mathrm{\textbf{r}},t;\mathrm{\textbf{r}}^\prime,t^\prime)=0,\quad\mbox{when
} t<t^\prime. \label{TDcausality}
\end{equation} Then, the space-FSE~(\ref{TDfse1}) becomes an integral equation,
\begin{equation}
\psi(\mathrm{\textbf{r}},t)=\psi_0(\mathrm{\textbf{r}},t)+\iint
G(\mathrm{\textbf{r}},t;\mathrm{\textbf{r}}^\prime,t^\prime)V(\mathrm{\textbf{r}}^\prime,t^\prime)\psi(\mathrm{\textbf{r}}^\prime,t^\prime)
\mathrm{d}^2\mathrm{\textbf{r}}^\prime\mathrm{d}t^\prime,\label{TDintegralform}
\end{equation} in which $\psi_0(\mathrm{\textbf{r}},t)$ satisfies the free-particle
Schr\"odinger equation,
\begin{equation}
[i\hslash\frac{\partial}{\partial t}+D_{\alpha}
(\hslash\nabla)^{\alpha}]\psi_0(\mathrm{\textbf{r}},t)=0.\label{TDfreeparticle}
\end{equation}By use of the method of separation of variables, the basic solution
to Eq.~(\ref{TDfreeparticle}) can be easily obtained.
\begin{equation}
\psi_0(\mathrm{\textbf{r}},t)=\mbox{e}^{i(\mathrm{\textbf{k}}\cdot\mathrm{\textbf{r}}-Et)/\hslash},\text{
{} {} (a constant product factor is omitted )}\label{TDfreeparticle1}
\end{equation}
where $E$ denotes the energy of the free particle, and
$\mathrm{\textbf{k}}=(k_x,k_y), $ in which $k_x,k_y$ are arbitrary
constants but satisfying
$|\mathrm{\textbf{k}}|=\sqrt{k_x^2+k_y^2}=(E/D_{\alpha})^{1/\alpha}$.
Replacing $\mathrm{\textbf{k}}$ by momentum $\mathrm{\textbf{p}}$,
and $E$ by $D_\alpha|\mathrm{\textbf{p}}|^\alpha$ respectively, the
free particle solution $\psi_0(\mathrm{\textbf{r}},t)$ can be
changed to the fractional plane wave solution \cite{laskin4},
\begin{equation}\psi_0(\mathrm{\textbf{r}},t)=\mbox{e}^{i(\mathrm{\textbf{p}}\cdot\mathrm{\textbf{r}}-D_{\alpha}|\mathrm{\textbf{p}}|^\alpha
t)/\hslash}.\label{TDfracplane}
\end{equation}
\\ Now we turn back to solve Eq.~(\ref{TDgreen}).  Defining the
Fourier transform pair, with respect to $\mathrm{\textbf{r}}$ and
$t$, of the Green's function
$G(\mathrm{\textbf{r}},t;\mathrm{\textbf{r}}^\prime,t^\prime)$ as
\begin{align}
&\hat{G}(\mathrm{\textbf{p}},\omega;\mathrm{\textbf{r}}^\prime,t^\prime)
=\int\mathrm{d}^2\mathrm{\textbf{r}}\int\mathrm{d}t\text{e}^{-i\mathrm{\textbf{p}}\cdot\mathrm{\textbf{r}}/\hslash-i\omega
t}G(\mathrm{\textbf{r}},t;\mathrm{\textbf{r}}^\prime,t^\prime),\label{TDfourier1}
\\&G(\mathrm{\textbf{r}},t;\mathrm{\textbf{r}}^\prime,t^\prime)
=\int\frac{\mathrm{d}^2\mathrm{\textbf{p}}}{(2\pi\hslash)^2}\int\frac{\mathrm{d}\omega}{2\pi}\text{e}^{i\mathrm{\textbf{p}}\cdot\mathrm{\textbf{r}}/\hslash+i\omega
t}\hat{G}(\mathrm{\textbf{p}},\omega;\mathrm{\textbf{r}}^\prime,t^\prime).\label{TDfourier2}
\end{align}
After taking Fourier transform, Eq.~(\ref{TDgreen}) can be changed to
\begin{equation}
-[\hslash\omega+D_{\alpha}|\mathrm{\textbf{p}}|^\alpha]\hat{G}(\mathrm{\textbf{p}},\omega;\mathrm{\textbf{r}}^\prime,t^\prime)
=\text{e}^{-i\mathrm{\textbf{p}}\cdot\mathrm{\textbf{r}^\prime}/\hslash-i\omega
t^\prime},
\end{equation}
That is, \begin{equation}
\hat{G}(\mathrm{\textbf{p}},\omega;\mathrm{\textbf{r}}^\prime,t^\prime)=-\frac{\text{e}^{-i\mathrm{\textbf{p}}\cdot\mathrm{\textbf{r}^\prime}/\hslash}\text{e}^{-i\omega
t^\prime}}{\hslash\omega+D_{\alpha}|\mathrm{\textbf{p}}|^\alpha}.
\end{equation}
Inverting the Fourier transform gives
\begin{equation}
G(\mathrm{\textbf{r}},t;\mathrm{\textbf{r}}^\prime,t^\prime)=-\int\frac{\mathrm{d}^2\mathrm{\textbf{p}}}{(2\pi\hslash)^2}\int\frac{\mathrm{d}\omega}{2\pi}
\frac{\text{e}^{i\mathrm{\textbf{p}}\cdot(\mathrm{\textbf{r}}-\mathrm{\textbf{r}^\prime})/\hslash}\text{e}^{i\omega
(t-t^\prime)}}{\hslash\omega+D_{\alpha}|\mathrm{\textbf{p}}|^\alpha}.\label{TDgreenint}
\end{equation} To calculate the integrals in the above formula, more
work is needed.
\begin{figure}[h]
 \centering
 \subfigure[The contour used for $t<t^\prime$. No poles is surrounded by the
contour. ]{ \label{TDfig1}\includegraphics[width=6cm]{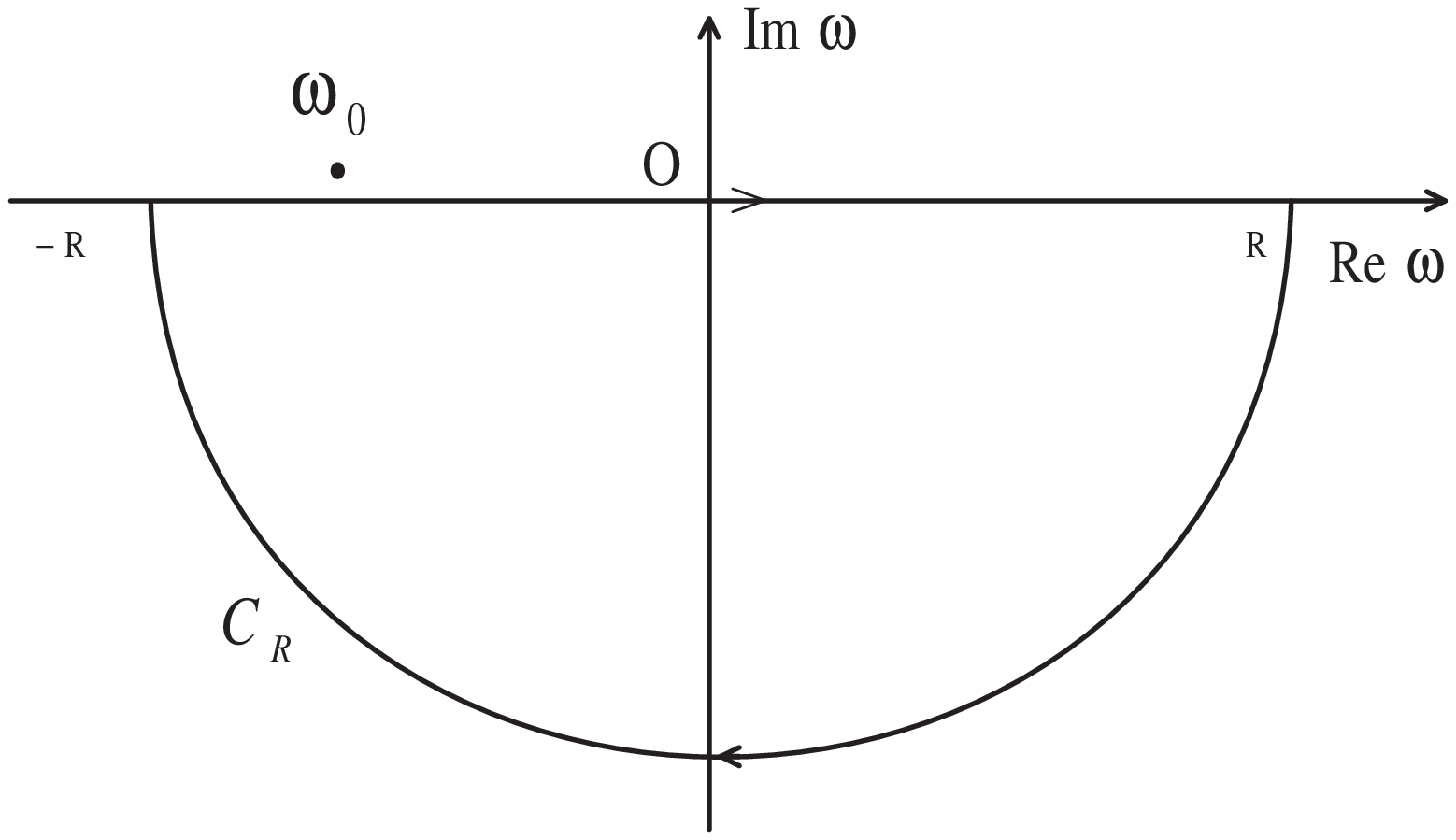}} \quad
\subfigure[The contour used for $t>t^\prime$.
  Only one pole $\omega_0$ is surrounded by the
  contour.]{ \label{TDfig2}\includegraphics[width=6cm]{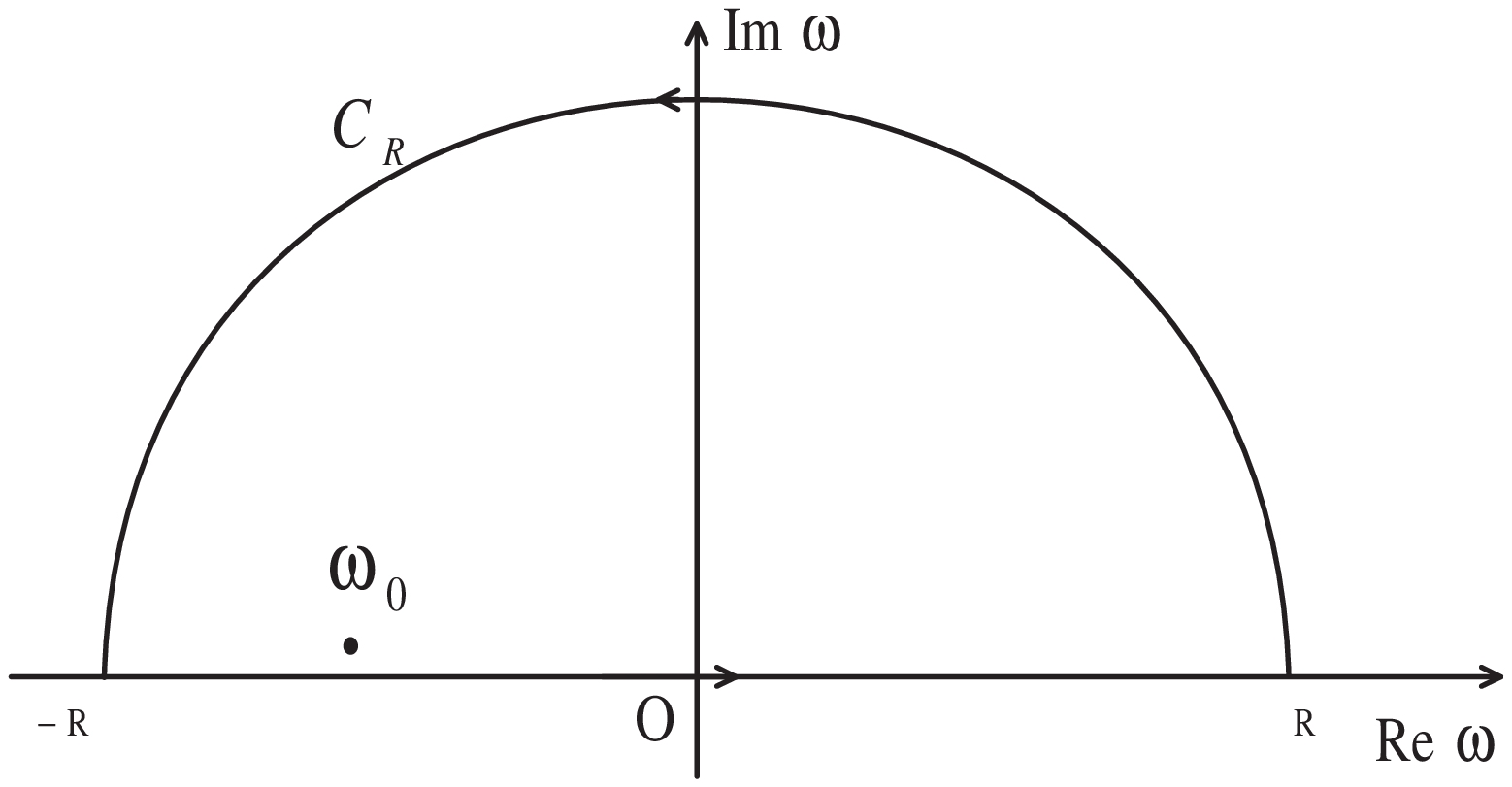}}
  \caption{The contour used to calculate the integral in Eq.~(\ref{TDintcal1}). $C_R$ is a semi-circular with radius $R$ ($R\rightarrow+\infty$).}
\end{figure}
 Let us consider the $\omega$ integration in the
complex $\omega$ plane,
\begin{equation}
\int^{+\infty}_{-\infty}\frac{\text{e}^{i\omega
(t-t^\prime)}}{\hslash\omega+D_{\alpha}|\mathrm{\textbf{p}}|^\alpha}\mathrm{d}\omega.\label{TDintcal1}
\end{equation} For the integrand, there is a pole on the path of
integration at $\omega_0=-D_{\alpha}|\mathrm{\textbf{p}}|^\alpha/\hslash$.
Considering the causality condition (\ref{TDcausality}), this
integration should be changed into
\begin{equation}
\int^{+\infty}_{-\infty}\frac{\text{e}^{i\omega
(t-t^\prime)}}{\hslash\omega+D_{\alpha}|\mathrm{\textbf{p}}|^\alpha-i\varepsilon}\mathrm{d}\omega,\label{TDintcal10}
\end{equation} where $\varepsilon$ is a positive infinitesimal \cite{Masujima}. Then
the pole is removed to the upper half plane. For $t<t^\prime$, we
use the contour shown in Fig. \ref{TDfig1} to carry out the contour
integral. Because no singularities lies inside the contour, the
integral is zero, which means the causality condition is satisfied.
For $t>t^\prime$, when we close the contour in the upper half-plane,
as shown in Fig. \ref{TDfig2}, we get the contribution from the pole
at $\omega_0$, and the result is
\begin{equation}
\lim_{\varepsilon\rightarrow0^+}\int^{+\infty}_{-\infty}\frac{\text{e}^{i\omega
(t-t^\prime)}}{\hslash\omega+D_{\alpha}|\mathrm{\textbf{p}}|^\alpha-i\varepsilon}\mathrm{d}\omega=2\pi
i\cdot\mbox{Res}\left\{\frac{\text{e}^{i\omega
(t-t^\prime)}}{\hslash\omega+D_{\alpha}|\mathrm{\textbf{p}}|^\alpha},\omega_0\right\}=
\frac{2\pi
i}{\hslash}\text{e}^{-iD_{\alpha}|\mathrm{\textbf{p}}|^\alpha
(t-t^\prime)/\hslash}.\label{TDintcal11}
\end{equation}Here, $\mbox{Res}\left\{*,\omega_0\right\}$ denotes the residue \cite{markj} of $*$ at $\omega_0$. Now Eq.~(\ref{TDgreenint}) becomes
\begin{equation}
G(\mathrm{\textbf{r}},t;\mathrm{\textbf{r}}^\prime,t^\prime)=\frac{1}{(2\pi\hslash)^2\hslash
i}\int\text{e}^{i\mathrm{\textbf{p}}\cdot(\mathrm{\textbf{r}}-\mathrm{\textbf{r}^\prime})/\hslash}\text{e}^{-iD_{\alpha}|\mathrm{\textbf{p}}|^\alpha
(t-t^\prime)/\hslash}\mathrm{d}^2\mathrm{\textbf{p}},\quad
t>t^\prime.\label{TDgreenint1}
\end{equation}
To execute the above integration, we choose the polar coordinates
$(p,\theta)$, with the positive direction of the $p$-axis along
$\mathrm{\textbf{r}}-\mathrm{\textbf{r}}^\prime$. Then,
$\mathrm{\textbf{p}}\cdot(\mathrm{\textbf{r}}-\mathrm{\textbf{r}^\prime})=p|\mathrm{\textbf{r}}-\mathrm{\textbf{r}^\prime}|\cos\theta$,
in which $p$ and $|\mathrm{\textbf{r}}-\mathrm{\textbf{r}}^\prime|$
denote the magnitudes of the vectors $\mathrm{\textbf{p}}$ and
$\mathrm{\textbf{r}}-\mathrm{\textbf{r}}^\prime$, respectively.
Thus, Eq.~(\ref{TDgreenint1}) is converted into
\begin{equation}
\begin{split}
G(\mathrm{\textbf{r}},t;\mathrm{\textbf{r}}^\prime,t^\prime)=\frac{1}{(2\pi\hslash)^2\hslash
i}\int^{2\pi} _{0}\mathrm{d}\theta\int^{+\infty}
_{0}\text{e}^{ip|\mathrm{\textbf{r}}-\mathrm{\textbf{r}^\prime}|\cos\theta/\hslash}\text{e}^{-iD_{\alpha}p^\alpha
(t-t^\prime)/\hslash}p\mathrm{d}p. \label{TDgreenint2}
\end{split}
\end{equation}Calculating the integral for $\theta$ yields
\begin{equation}
\int^{2\pi}
_{0}\text{e}^{ip|\mathrm{\textbf{r}}-\mathrm{\textbf{r}^\prime}|\cos\theta/\hslash}\mathrm{d}\theta=2\int^{\pi}
_{0}\cos({ip|\mathrm{\textbf{r}}-\mathrm{\textbf{r}^\prime}|\cos\theta/\hslash})\mathrm{d}\theta=2\pi
J_0(p|\mathrm{\textbf{r}}-\mathrm{\textbf{r}^\prime}|/\hslash),\label{TDintcalcu1}
\end{equation}where $J_0(r)$ is the Bessel function of the first kind of order
$0$. Furthermore, taking into account the series form of the Bessel
function of the first kind of order $\nu$, that is,
\begin{equation}
\begin{split}
J_\nu(z)=\frac{z^\nu}{2^\nu}\sum^{\infty}_{k=0}\frac{(-1)^{k}z^{2k}}{2^{2k}k!\Gamma\left(\nu+k+1\right)},
\end{split}
\end{equation}
we obtain
\begin{equation}
\begin{split}
G(\mathrm{\textbf{r}},t;\mathrm{\textbf{r}}^\prime,t^\prime)=\frac{2\pi}{(2\pi\hslash)^2\hslash
i}\sum^{\infty}_{k=0}\frac{(-1)^{k}(|\mathrm{\textbf{r}}-\mathrm{\textbf{r}^\prime}|/\hslash)^{2k}}{2^{2k}(k!)^2}
 I_k, \label{TDgreenseries1}
\end{split}
\end{equation}
in which \begin{equation} I_k=\int^{+\infty}
_{0}p^{2k+1}\text{e}^{-iD_{\alpha}p^\alpha
(t-t^\prime)/\hslash}\mathrm{d}p.\label{TDgreenseriesik}
\end{equation}Making substitution $\tilde{p}=[D_{\alpha}(t-t^\prime)/\hslash]^{1/\alpha}p\equiv\xi\cdot
p$ gives
\begin{equation}
I_k=\xi^{-(2k+2)}\int^{+\infty} _{0}p^{2k+1}\text{e}^{-i
p^\alpha}\mathrm{d}p,\label{TDikcal1}
\end{equation}in which $\tilde{p}$ has been replaced by $p$ for
simplicity. \begin{figure}[h]
   \centering
 \includegraphics[width=6cm]{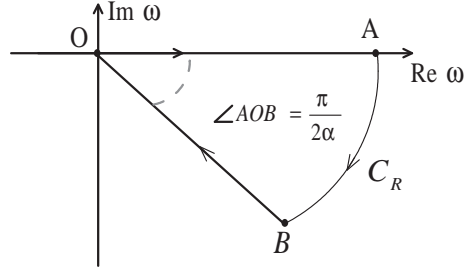}
 \caption{The contour used to calculate the integral in
 Eq.~(\ref{TDikcal1}).}
 \label{TDfig3}
\end{figure}Then using the contour in Fig. \ref{TDfig3}, the above integral can be
calculated.
\begin{equation}
\int^{+\infty}_{0}p^{2k+1}\text{e}^{-ip^\alpha}\mathrm{d}p=\frac1\alpha\Gamma(\frac{2k+2}{\alpha})\mbox{e}^{-\frac{k+1}{\alpha}\pi
i}.\label{TDikcal2}
\end{equation} Finally, we obtain the Green's function for the fractional Schr\"odinger equation in a series form,
\begin{equation}
G(\mathrm{\textbf{r}},t;\mathrm{\textbf{r}}^\prime,t^\prime)=\frac{\xi^{-2}}{2\alpha\pi\hslash^3i}
\sum^{\infty}_{k=0}\frac{(-1)^{k}}{(k!)^2}\left[\frac{|\mathrm{\textbf{r}}-\mathrm{\textbf{r}^\prime}|}{2\xi\hslash}\right]^{2k}
 \Gamma(\frac{2k+2}{\alpha})\mbox{e}^{-\frac{k+1}{\alpha}\pi
i},\label{TDgreenintimefinal}
\end{equation} in which $\xi=[D_{\alpha}(t-t^\prime)/\hslash]^{1/\alpha}$.
It can be proved that this Green's function reduces to the one in the standard quantum mechanics when
$\alpha=2$.

\section{ {Time-independent Case}}
\label{sec2}
\begin{figure}[ht]
 \includegraphics[width=8cm]{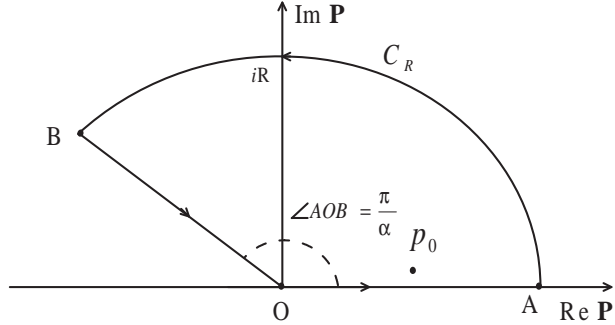}
 \centering
 \caption{The contour used to calculate $I_1(r)$.
 $C_R=\stackrel{\frown}{AB}$ is a circular segment with radius $R$.
 Only one pole $p_0$ is surrounded by the contour.
 }
 \label{fig1}
 \end{figure}

When the potential is independent of $t$, that is,
$V(\mathrm{\textbf{r}},t)=V(\mathrm{\textbf{r}})$, after separation
of variables, the steady state form of the fractional Schr\"odinger
equation (\ref{TDfse1}) can be obtained \cite{dong3}:
\begin{equation}
-D_{\alpha}
(\hslash\nabla)^{\alpha}\phi(\mathrm{\textbf{r}})+V(\mathrm{\textbf{r}})\phi(\mathrm{\textbf{r}})=E\phi(\mathrm{\textbf{r}}),\label{fse3}
\end{equation}
where $\phi(\mathrm{\textbf{r}})$ is related to
$\psi(\mathrm{\textbf{r}},t)$ by
$\psi(\mathrm{\textbf{r}},t)=\phi(\mathrm{\textbf{r}})\mbox{e}^{-iEt/\hslash}$
, in which $E$ denotes the energy of the quantum system.
\\Eq.~(\ref{fse3}) can be rewritten as
\begin{equation}
[(\hslash\nabla)^{\alpha}+\kappaup]\phi(\mathrm{\textbf{r}})=Q,\label{fse4}
\end{equation}
where $\kappaup=E/D_{\alpha}$,$Q=V\phi/D_{\alpha}$. Now, we define
the \textbf{Green's function} of the above equation by
\begin{equation}
[(\hslash\nabla)^{\alpha}+\kappaup]G(\mathrm{\textbf{r}})=\delta^2(\textbf{r}),\label{green}
\end{equation}
then $\phi$ can be expressed as:
\begin{equation}
\phi(\mathrm{\textbf{r}})=\phi_0(\mathrm{\textbf{r}})+\int
G(\mathrm{\textbf{r}}-\mathrm{\textbf{r}}_0)Q(\mathrm{\textbf{r}}_0)\mathrm{d}^2\mathrm{\textbf{r}}_0,\label{green1}
\end{equation} where $\phi_0(\mathrm{\textbf{r}})$ satisfies the free-particle
Schr\"odinger equation,
\begin{equation}
[(\hslash\nabla)^{\alpha}+\kappaup]\phi_0(\mathrm{\textbf{r}})=0.\label{free}
\end{equation}It is easy to prove that the basic solution
to Eq.~(\ref{free}) is
\begin{equation}
\phi_0(\mathrm{\textbf{r}})=e^{i\mathrm{\textbf{k}}\cdot\mathrm{\textbf{r}}},\mbox{
{} {} (a constant product factor is omitted )}\label{freeparticle}
\end{equation}
where  $\mathrm{\textbf{k}}=(K_x,K_y), $ and
$|\mathrm{\textbf{k}}|=\sqrt{K_x^2+K_y^2}=\kappaup^{1/\alpha}/\hslash$.
\\ Now we turn back to solve Eq.~(\ref{green}).  Defining
\begin{equation}\hat{G}(\mathrm{\textbf{p}})=\int\mbox{e}^{-i\mathrm{\textbf{p}}\cdot\mathrm{\textbf{r}}/\hslash}G(\mathrm{\textbf{r}})\mathrm{d}^2\mathrm{\textbf{r}},
\mbox{ and }
G(\mathrm{\textbf{r}})=\frac1{(2\pi\hslash)^2}\int\mbox{e}^{i\mathrm{\textbf{p}}\cdot\mathrm{\textbf{r}}/\hslash}\hat{G}(\mathrm{\textbf{p}})\mathrm{d}^2\mathrm{\textbf{p}},
\end{equation}
after taking the Fourier transform, Eq.~(\ref{green}) is changed
into
\begin{equation}
[\kappaup-|\mathrm{\textbf{p}}|^\alpha]\hat{G}(\mathrm{\textbf{p}})=1,
\end{equation}
That is, \begin{equation}
\hat{G}(\mathrm{\textbf{p}})=\frac{1}{\kappaup-|\mathrm{\textbf{p}}|^\alpha}.
\end{equation}
Inverting the Fourier transform gives
\begin{equation}
G(\mathrm{\textbf{r}})=\frac1{(2\pi\hslash)^2}\int\frac{\mbox{e}^{i\mathrm{\textbf{p}}\cdot\mathrm{\textbf{r}}/\hslash}}{\kappaup-|\mathrm{\textbf{p}}|^\alpha}\mathrm{d}^2\mathrm{\textbf{p}}.\label{green2}
\end{equation}
To calculate the integral in the above formula, more work is needed.
Denoting the magnitude of the vectors $\mathrm{\textbf{p}}$ and
$\mathrm{\textbf{r}}$ by $p$ and $r$, respectively, with the help of
the polar coordinates $(p,\theta)$, Eq.~(\ref{green2}) can be
converted into
\begin{equation}
G(\mathrm{\textbf{r}})=\frac1{(2\pi\hslash)^2}\int^{2\pi}
_{0}\mathrm{d}\theta\int^{+\infty}
_{0}\frac{\mbox{e}^{ipr\cos\theta/\hslash}}{\kappaup-p^\alpha}p\mathrm{d}p=\frac{1}{(\pi\hslash)^2}\int^{\pi/2}
_{0}\mathrm{d}\theta\int^{+\infty}
_{0}\frac{\cos({pr\cos\theta/\hslash})}{\kappaup-p^\alpha}p\mathrm{d}p.\label{origingreen}
\end{equation}
When considering the quantum scattering problems, the energy of the
quantum system satisfies $E>0$,  that is, $\kappaup>0$.  To evaluate
the integral in Eq.~(\ref{origingreen}), we should consider the
quantum mechanics effect, and define two integrals, for the outgoing
wave and the incoming wave respectively, as follows:
\begin{equation}
G^\pm(\mathrm{\textbf{r}})=\frac{1}{(\pi\hslash)^2}\int^{\pi/2}
_{0}\mathrm{d}\theta\int^{+\infty}
_{0}\frac{\cos({pr\cos\theta/\hslash})}{\kappaup-p^\alpha\pm
i\varepsilon}p\mathrm{d}p,\label{origingreen1}
\end{equation}
where $\varepsilon$ is a positive infinitesimal (This manner can be
called the \textquoteleft\textquoteleft $i\varepsilon$
prescription\textquoteright\textquoteright, see \cite{Shankar}).
Note that in the quantum scattering problems, one is usually
interested in the outgoing wave with $G^+(\mathrm{\textbf{r}})$
being considered. From Eqs.~(\ref{origingreen1}), we know
$G^-(\mathrm{\textbf{r}})=[G^+(\mathrm{\textbf{r}})]^*$, and
$G(\mathrm{\textbf{r}})=[G^+(\mathrm{\textbf{r}})+G^-(\mathrm{\textbf{r}})]/2$,
so we only need to calculate $G^+(\mathrm{\textbf{r}})$ here.
Substituting $p$ for $p/\kappaup^{1/\alpha}$ in
Eq.~(\ref{origingreen1}) yields
\begin{equation}
G^+(\mathrm{\textbf{r}})=\frac{1}{(\pi\hslash)^2}\int^{\pi/2}
_{0}\mathrm{d}\theta\int^{+\infty}
_{0}\frac{\cos({p\kappaup^{1/\alpha}r\cos\theta/\hslash})}{1-p^\alpha+
i\eta}p\mathrm{d}p,\label{origingreen2}
\end{equation}
where $\eta=\varepsilon/\kappaup$. Moreover,
$G^+(\mathrm{\textbf{r}})$ can be rewritten as
\begin{equation}
G^+(\mathrm{\textbf{r}})=\frac{\kappaup^{(2-\alpha)/\alpha}}{2(\pi\hslash)^2}[I_1(r\kappaup^{1/\alpha}/\hslash)+I_2(r\kappaup^{1/\alpha}/\hslash)],\label{greenforoutwave1}
\end{equation} in which the two functions $I_1(r)$ and $I_1(r)$ are
defined as
\begin{equation}
I_1(r)=\int^{\pi/2} _{0}\mathrm{d}\theta\int^{+\infty}
_{0}\frac{pe^{ipr\cos\theta}}{1-p^\alpha+i\eta}\mathrm{d}p,\label{greenforoutwave1i1}
\end{equation} and
\begin{equation}
I_2(r)=\int^{\pi/2} _{0}\mathrm{d}\theta\int^{+\infty}
_{0}\frac{pe^{-ipr\cos\theta}}{1-p^\alpha+i\eta}\mathrm{d}p.\label{greenforoutwave1i2}
\end{equation}
The integral for $p$ can be evaluated by. To calculate the two
integral, we can make use of the contour integral method and the
residue theorem. It should be noted that all of the integrals in
this paper take the Cauchy principal values. For $I_1(r)$, the
contour is chosen as shown in Fig.~\ref{fig1}.
 Note that in the
definition of $G^+(\mathrm{\textbf{r}})$ in
Eq.~(\ref{origingreen1}), the pole has been removed from the real
axis to the upper half plane. There is one pole to be considered,
$p_0=r_\eta e^{i\theta_\eta}$, in which $r_\eta, \theta_\eta>0$, and
$r_\eta\rightarrow1$, $\theta_\eta\rightarrow0$ as
$\eta\rightarrow0^+$, and the expression of $r_\eta$ and
$\theta_\eta$ needs not be specified here. On the circular segment
$C_R$, the formula
$$\lim\limits_{R\rightarrow+\infty}\frac{p}{1-p^\alpha}=0,(1<\alpha\leq2)$$ holds uniformly, so according
to the Jordan's lemma \cite{markj}, the integral
$$\int_{C_R}\frac{p\mbox{e}^{ipr\cos\theta}}{1-p^\alpha}\mathrm{d}p$$ is
zero when $R\rightarrow+\infty$.  Therefore, we have
\begin{align}
I_1(r)&=\lim\limits_{\eta\rightarrow0^+}\lim\limits_{R\rightarrow+\infty}\left\{2\pi
i\cdot\mbox{Res}\Big(\frac{pe^{ipr\cos\theta}}{1-p^\alpha+i\eta};p_{0}\Big)-\Bigg(\int_{C_R}+\int_{\overrightarrow{BO}}\Bigg)\frac{pe^{ipr\cos\theta}}{1-p^\alpha+i\eta}\mathrm{d}p\right\}
\nonumber\\&=-\frac{2\pi
i}{\alpha}e^{ir\cos\theta}+I_{\overrightarrow{OB}},\label{I1calculate}
\end{align}where
\begin{equation}
I_{\overrightarrow{OB}}=\lim\limits_{R\rightarrow+\infty}\int_{\overrightarrow{OB}}\frac{pe^{ipr\cos\theta}}{1-p^\alpha}\mathrm{d}p.\label{IOB}
\end{equation}
Taking into account that $p=\tilde{p}e^{\frac{\pi i}{\alpha}}$
($\tilde{p}\geq0$) on $\overrightarrow{OB}$, inserting this
expression for $p$ into Eq.~(\ref{IOB}) gives
\begin{equation}
I_{\overrightarrow{OB}}=e^{\frac{2\pi
i}{\alpha}}[\mathbb{I}_1(r,\theta)+i\cdot
\mathbb{I}_2(r,\theta)],\label{IOBcal}
\end{equation} where
\begin{align}
&\mathbb{I}_1(r,\theta)=\int^{+\infty}
_{0}\frac{p}{1+p^\alpha}e^{-\sin(\pi/\alpha)pr\cos\theta}\cos[\cos(\pi/\alpha)pr\cos\theta]\mathrm{d}p,\label{IOBcal1}
\\&\mathbb{I}_2(r,\theta)=\int^{+\infty}
_{0}\frac{p}{1+p^\alpha}e^{-\sin(\pi/\alpha)pr\cos\theta}\sin[\cos(\pi/\alpha)pr\cos\theta]\mathrm{d}p,\label{IOBcal2}
\end{align}in which $\tilde{p}$ has been replaced by $p$ for
simplicity. Therefore, we get
\begin{align}
I_1(r)&=\int^{\pi/2} _{0}\left\{-\frac{2\pi
i}{\alpha}e^{ir\cos\theta}+e^{\frac{2\pi
i}{\alpha}}[\mathbb{I}_1(r,\theta)+i\cdot
\mathbb{I}_2(r,\theta)]\right\}\mathrm{d}\theta
\\&=-\frac{2\pi
i}{\alpha}\int^{\pi/2}
_{0}e^{ir\cos\theta}\mathrm{d}\theta+e^{\frac{2\pi
i}{\alpha}}[\mathcal {I}_1(r)+i\mathcal {I}_2(r)],\label{i1calcu1}
\end{align}where \begin{align}\mathcal {I}_1(r)=\int^{\pi/2}
_{0}\mathbb{I}_1(r,\theta)\mathrm{d}\theta, \mbox{ and } \mathcal
{I}_2(r)=\int^{\pi/2} _{0}\mathbb{I}_2(r,\theta)\mathrm{d}\theta.\label{twointegral}
\end{align}
In a similar way, $I_2(r)$ can also be calculated, and expressed in
terms of $\mathcal {I}_1(r)$ and $\mathcal {I}_2(r)$. Note that
there is no poles to be considered when calculating $I_2(r)$. The
result is
\begin{align}
I_2(r)=e^{-\frac{2\pi i}{\alpha}}[\mathcal {I}_1(r)-i\mathcal
{I}_2(r)].\label{i2calcu1}
\end{align}
The first integral in Eq.~(\ref{i1calcu1}) can be calculated as
\begin{align} \int^{\pi/2}
_{0}e^{ir\cos\theta}\mathrm{d}\theta
=&\int^{\pi/2}
_{0}\cos(r\cos\theta)\mathrm{d}\theta+i\int^{\pi/2}
_{0}\sin(r\cos\theta)\mathrm{d}\theta
\\=&\frac{\pi}{2}\left[J_0(r)+iH_0(r)\right],\label{intini1}
\end{align}where $J_0(r)$ is the Bessel function of the first kind of order $0$, and
$H_0(r)$ is the Struve function of order $0$.
The two integrals $\mathcal {I}_1(r)$ and $\mathcal {I}_2(r)$ in
Eq.~(\ref{twointegral}) can be expressed in terms of Fox's H-function,
using the Mellin transform and its inverse transform. With the help
of the following formulas for the Mellin transform (see P.1189 of
\cite{Polyanin})
\begin{align} &\mathcal
{M}\mbox{
}\Big\{e^{-ax}\sin(bx),s\Big\}=\frac{\Gamma(s)\sin[s\arctan\cot(b/a)]}{(a^2+b^2)^{s/2}},
\quad a>0,\mbox{ Re }s>-1
\\&\mathcal
{M}\mbox{
}\Big\{e^{-ax}\cos(bx),s\Big\}=\frac{\Gamma(s)\cos[s\arctan\cot(b/a)]}{(a^2+b^2)^{s/2}},\quad
a>0,\mbox{ Re }s>0
\end{align} and considering the formula (see formula 3.241.2 of \cite{Gradsh}),
\begin{equation}
\int^{+\infty}
_{0}\frac{x^{a-1}}{x^b+1}\mathrm{d}x=\frac{\pi}{b}\csc(\frac{a}{b}\pi),\qquad
\mbox{ Re }b>\mbox{ Re }a>0\label{formula1}
\end{equation}the Mellin transform of $\mathcal {I}_1(r)$ and $\mathcal {I}_2(r)$ can be calculated as
\begin{align}
\tilde{\mathcal {I}}_1(s)&=\mathcal {M}\mbox{ }\Big\{\mathcal
{I}_1(r),s\Big\}=\int^{\pi/2} _{0}\mathcal {M}\mbox{
}\Big\{\mathbb{I}_1(r,\theta),s\Big\}\mathrm{d}\theta
\\&=\Gamma(s)\cos(\lambda\pi s)\cdot\int^{\pi/2}
_{0}(\cos\theta)^{-s}\mathrm{d}\theta\cdot\int^{+\infty}
_{0}\frac{p^{1-s}}{1+p^\alpha}\mathrm{d}p
\\&=\frac{\pi}{\alpha}\frac{\Gamma(s)\cos(\lambda\pi
s)}{2^{s+1}\sin(\frac{2-s}{\alpha}\pi)}\beta\left(\frac{1-s}{2},\frac{1-s}{2}\right)=\frac{\pi}{\alpha}
\frac{\Gamma(s)\Gamma(1-\frac{2-s}{\alpha})\Gamma(\frac{1-s}{2})\Gamma(\frac{1-s}{2})\Gamma(\frac{2-s}{\alpha})}
{2^{s+1}\Gamma(\frac12+\lambda s)\Gamma(1-s)\Gamma(\frac12-\lambda
s)},\nonumber
\end{align}
\begin{align}
\tilde{\mathcal {I}}_2(s)=\mathcal {M}\mbox{ }\Big\{\mathcal
{I}_2(r),s\Big\}=\int^{\pi/2} _{0}\mathcal {M}\mbox{
}\Big\{\mathbb{I}_2(r,\theta),s\Big\}\mathrm{d}\theta=-\frac{\pi}{\alpha}\frac{\Gamma(s)\Gamma(1-\frac{2-s}{\alpha})\Gamma(\frac{1-s}{2})\Gamma(\frac{1-s}{2})\Gamma(\frac{2-s}{\alpha})}
{2^{s+1}\Gamma(\lambda s)\Gamma(1-s)\Gamma(1-\lambda
s)},\label{mellinfori2}
\end{align}where $\lambda=-\arctan[\cot(\pi/\alpha)]/\pi$, and
$0\leq\lambda<1/2$ when $1<\alpha\leq2$. Additionally, $\lambda=0$
only when $\alpha=2$, and at that time, $\tilde{\mathcal
{I}_2}(s)\equiv0$, so that $\mathcal{I}_2(r)\equiv0$. Inverting the
Mellin transform, and comparing the expression with the definition
of the Fox's H-function, we obtain
\begin{align}
\mathcal{I}_1(r)=\frac{1}{2\pi i}\int^{c+i\infty}
_{c-i\infty}r^{-s}\tilde{\mathcal{I}}_1(s)\mathrm{d}s=\frac{\pi}{2\alpha}H^{2,3}_{4,4}\left[2r\mbox{
}\Bigg| \begin{aligned}&{\mbox{
}(1/2,1/2),(1/2,1/2),(1-2/\alpha,1/\alpha),(1/2,\lambda)}
\\&{\mbox{
}(0,1),(1-2/\alpha,1/\alpha),(0,1),(1/2,\lambda)}
\end{aligned}\right]\equiv\frac{\pi}{2\alpha}\mathscr{H}_1(r),\label{Icalfinal1}
\end{align}
\begin{align}
\mathcal{I}_2(r)=\frac{1}{2\pi i}\int^{c+i\infty}_{c-i\infty}r^{-s}\tilde{\mathcal{I}}_2(s)\mathrm{d}s=-\frac{\pi}{2\alpha}H^{2,3}_{4,4}\left[2r\mbox{
}\Bigg| \begin{aligned}&{\mbox{
}(1/2,1/2),(1/2,1/2),(1-2/\alpha,1/\alpha),(0,\lambda)}
\\&{\mbox{
}(0,1),(1-2/\alpha,1/\alpha),(0,1),(0,\lambda)}
\end{aligned}\right]\equiv-\frac{\pi}{2\alpha}\mathscr{H}_2(r).\label{Icalfinal2}
\end{align}
Finally, using the results in Eqs.~(\ref{i1calcu1}),
(\ref{i2calcu1}), (\ref{Icalfinal1})  and (\ref{Icalfinal2}), from
Eq.~(\ref{greenforoutwave1}), the Green's function $G^+(r)$ for the
outgoing wave can be calculated in terms of the Bessel function, Struve function and
Fox's H-function as:
\begin{align}
G^+(\mathrm{\textbf{r}})=\frac{\kappaup^{(2-\alpha)/\alpha}}{2\alpha\hslash^2i}\left[J_0(r\kappaup^{1/\alpha}/\hslash)+iH_0(r\kappaup^{1/\alpha}/\hslash)\right]+\frac{\kappaup^{(2-\alpha)/\alpha}}{2\alpha\pi\hslash^2}\mathcal
{G}(r\kappaup^{1/\alpha}/\hslash), \label{grforoutfinal}
\end{align}where
\begin{equation}
\mathcal
{G}(r)=\cos{\frac{2\pi}{\alpha}}\mathscr{H}_1(r)+\sin{\frac{2\pi}{\alpha}}\mathscr{H}_2(r).\label{mathcalgrfori1i2}
\end{equation}

\section{ {Asymptotic properties of the Green's functions and Applications to the Scattering Problems}}
\label{sec3}In the scattering problems, we usually consider the
behavior of the particles far away from the scattering center, and
assume the potential $V(\mathrm{\textbf{r}})$ is non-zero only in a
small domain. In this section, we discuss the asymptotic properties
of the Green's functions obtained in the previous sections, and then apply them to study the fractional scattering problems.
\subsection{ {Time-dependent case]}}
To get the asymptotic formula for the Green's function, we can rewritten it in terms of the H-function. The formula of the Green's function can be rewritten as,
\begin{equation}
\begin{split}
G(\mathrm{\textbf{r}},t;\mathrm{\textbf{r}}^\prime,t^\prime)&=\frac{\xi^{-2}}{2\alpha\pi\hslash^3i}
\sum^{\infty}_{k=0}\frac{(-1)^{k}}{(k!)^2}\left(\frac{|\mathrm{\textbf{r}}-\mathrm{\textbf{r}^\prime}|}{2\xi\hslash}\right)^{2k}
 \Gamma(\frac{2k+2}{\alpha})\left(\cos\frac{k+1}{\alpha}\pi-i\sin\frac{k+1}{\alpha}\pi\right)
 \\&=\frac{\xi^{-2}}{2\alpha\hslash^3i}
\sum^{\infty}_{k=0}\left[\frac{\Gamma(\frac{2k+2}{\alpha})}{\Gamma(k+1)\Gamma(\frac12+\frac1\alpha+\frac k\alpha)\Gamma(\frac12-\frac1\alpha-\frac k\alpha)}
-i\frac{\Gamma(\frac{2k+2}{\alpha})}{\Gamma(k+1)\Gamma(\frac1\alpha+\frac k\alpha)\Gamma(1-\frac1\alpha-\frac k\alpha)}\right]\frac{(-1)^{k}}{k!}y^k,\label{greentohf}
\end{split}
\end{equation}
 in which $y=\left(\frac{|\mathrm{\textbf{r}}-\mathrm{\textbf{r}^\prime}|}{2\xi\hslash}\right)^{2}$. Comparing this equation to the definition of Fox's H-function, we can get that
 \begin{equation}
G(\mathrm{\textbf{r}},t;\mathrm{\textbf{r}}^\prime,t^\prime)=\frac{\xi^{-2}}{2\alpha\hslash^3i}\left[\mathcal {H}_1(y)-i\mathcal {H}_2(y)\right]
\end{equation} in which
 \begin{align}
&\mathcal {H}_1(y)=H^{1,1}_{2,3}\left[y\mbox{
}\Bigg| \begin{aligned}&{\mbox{
}(1-2/\alpha,2/\alpha),(1/2-1/\alpha,1/\alpha)}
\\&{\mbox{
}(0,1),(0,1),(1/2-1/\alpha,1/\alpha)}
\end{aligned}\right],\label{greentohf1}
\\&\mathcal {H}_2(y)=H^{1,1}_{2,3}\left[y\mbox{
}\Bigg| \begin{aligned}&{\mbox{
}(1-2/\alpha,2/\alpha),(1-1/\alpha,1/\alpha)}
\\&{\mbox{
}(0,1),(0,1),(1-1/\alpha,1/\alpha)}
\end{aligned}\right].\label{greentohf2}
\end{align}
By use the asymptotic properties of H function, we can get
 \begin{equation}
 \begin{split}
 G(\mathrm{\textbf{r}},t;\mathrm{\textbf{r}}^\prime,t^\prime)&=2\pi iAy^{(\mu+1/2)/\Delta}\mbox{e}^{i(B+Cy^{1/\Delta}-\pi/\alpha)}+\mbox{o}(y^{(\mu+1/2)/\Delta}).
 \\&=-\frac{|\mathrm{\textbf{r}}-\mathrm{\textbf{r}}^\prime|^{(2-\alpha)/(\alpha-1)}}{2\pi\hslash^2[\alpha D_\alpha(t-t^\prime)]^{1/(\alpha-1)}\sqrt{\alpha-1}}\exp\left\{i\frac{\alpha-1}{\hslash}\left(\frac{|\mathrm{\textbf{r}}-\mathrm{\textbf{r}}^\prime|^\alpha}{\alpha^\alpha D_\alpha(t-t^\prime)}\right)^{\frac{1}{\alpha-1}}\right\}+\mbox{o}\left(|\mathrm{\textbf{r}}-\mathrm{\textbf{r}}^\prime|^{\frac{2-\alpha}{\alpha-1}}\right)
 \end{split}, \label{asmpforg}
 \end{equation}
 In the scattering problem, we can use merely the first term of the
asymptotic formula (\ref{asmpforg}) for
$G(\mathrm{\textbf{r}},t;\mathrm{\textbf{r}}^\prime,t^\prime)$, then
an approximate wave function for the scattering problem can be
obtained. Let's invoke the Born approximation \cite{grif,born}:
Suppose the incoming plane wave is not substantially altered by the
potential. Then, in Eq.~(\ref{TDintegralform}), it makes sense to use
\begin{equation}
\psi(\mathrm{\textbf{r}}^\prime,t^\prime)\approx\psi_0(\mathrm{\textbf{r}}^\prime,t^\prime)=e^{i(\mathrm{\textbf{k}}\cdot\mathrm{\textbf{r}}^\prime-Et^\prime)/\hslash}.\label{TDorigin2}
\end{equation} Then Eq.~(\ref{TDintegralform}) becomes
\begin{equation}
\begin{split}
\psi(\mathrm{\textbf{r}},t)\approx\psi_0&(\mathrm{\textbf{r}},t)-\frac{(\alpha D_\alpha)^{1/(1-\alpha)}}{2\pi\hslash^2\sqrt{\alpha-1}}\int
\frac{|\mathrm{\textbf{r}}-\mathrm{\textbf{r}^\prime}|^{(2-\alpha)/(\alpha-1)}}{(t-t^\prime)^{1/(\alpha-1)}}
\exp\Bigg\{i\frac{\alpha-1}{\hslash}\left(\frac{|\mathrm{\textbf{r}}-\mathrm{\textbf{r}}^\prime|^\alpha}{\alpha^\alpha D_\alpha(t-t^\prime)}\right)^{\frac{1}{\alpha-1}}
+\\&i(\mathrm{\textbf{k}}\cdot\mathrm{\textbf{r}}^\prime-Et^\prime)/\hslash\Bigg\}
V(\mathrm{\textbf{r}}^\prime,t^\prime)\mathrm{d}^2\mathrm{\textbf{r}}^\prime\mathrm{d}t^\prime.\label{TDapproxwave1}
\end{split}
\end{equation}Further more, assuming the potential $V(\mathrm{\textbf{r}})$ is non-zero only in a
small domain, then we have $|\mathrm{\textbf{r}}-\mathrm{\textbf{r}^\prime}|\approx|\mathrm{\textbf{r}}|=r$, we can get
\begin{equation}
\begin{split}
\psi(\mathrm{\textbf{r}},t)\approx\psi_0&(\mathrm{\textbf{r}},t)-\frac{(\alpha D_\alpha r^{\alpha-2})^{\frac{1}{1-\alpha}}}{2\pi\hslash^2\sqrt{\alpha-1}}\int
(t-t^\prime)^{\frac{1}{1-\alpha}}
\exp\Bigg\{i\frac{\alpha-1}{\hslash}\left(\frac{(r/\alpha)^\alpha}{D_\alpha(t-t^\prime)}\right)^{\frac{1}{\alpha-1}}
+i(\mathrm{\textbf{k}}\cdot\mathrm{\textbf{r}}^\prime-Et^\prime)/\hslash\Bigg\}
V(\mathrm{\textbf{r}}^\prime,t^\prime)\mathrm{d}^2\mathrm{\textbf{r}}^\prime\mathrm{d}t^\prime.\label{TDapproxwave2}
\end{split}
\end{equation}The second term of the right side of the above formula gives the
approximate scattering wave function.
\par We can also generate a
series of higher-order corrections to the approximate wave function.
From Eq.~(\ref{TDintegralform}), we can build an iteration scheme for
the wave function as
\begin{equation}
\psi^{(n)}(\mathrm{\textbf{r}},t)=\psi_0(\mathrm{\textbf{r}},t)+\int
G(\mathrm{\textbf{r}},t;\mathrm{\textbf{r}}^\prime,t^\prime)V(\mathrm{\textbf{r}}^\prime,t^\prime)\psi^{(n-1)}(\mathrm{\textbf{r}}^\prime,t^\prime)
\mathrm{d}^3\mathrm{\textbf{r}}^\prime\mathrm{d}t^\prime.\label{TDiteration}
\end{equation}Note that
$\phi^{(0)}(\mathrm{\textbf{r}},t)=\psi_0(\mathrm{\textbf{r}},t)=e^{i(\mathrm{\textbf{k}}\cdot\mathrm{\textbf{r}}-Et)/\hslash}$,
and $\phi^{(n)}$ is the $n$th-order corrections to the wave
function. For a given potential function, using
Eq.~(\ref{TDiteration}), the analytical approximate solutions of every
order can be obtained. In a series form, we have
\begin{equation}
\phi=\phi_0+\int GV\phi_0+\iint GVGV\phi_0+\iiint
GVGVGV\phi_0+\cdots.
\end{equation}In each integrand only the incident wave function ($\phi_0$)
appears, together with more and more powers of $GV$.
\subsection{ {Time-independent case}}
For the Green's function , for the the Bessel function and Struve function, we have the following  asymptotic formulas,
\begin{equation}
H_0[r]=Y_0(r)+\mbox{O} (|r|^{-1}),Y_0(r)\sim\sqrt{\frac{2}{\pi r}}\sin[r-\frac{\pi}{4}],\mbox{ { } }r\rightarrow\infty,
\label{}
\end{equation}
\begin{equation}
J_0[r]\sim\sqrt{\frac{2}{\pi r}}\cos[r-\frac{\pi}{4}],\mbox{ when}
 r\rightarrow+\infty.\label{}
\end{equation}   Meantime, with the help of the properties of the H function, we obtain  the
asymptotic formulas for the two H functions, $\mathscr{H}_1(r)$, and $\mathscr{H}_2(r)$,
\begin{equation}
\mathscr{H}_1(r),\mathscr{H}_2(r)=\mbox{o}\left(\frac{\log
(r)}{r}\right),\mbox{ when}
 r\rightarrow+\infty.\label{apph1}
\end{equation}
Therefore, the Green's function can be approximate to
\begin{equation}
\begin{split}
G^+(\mathrm{\textbf{r}})&\sim\frac{\kappaup^{(2-\alpha)/\alpha}}{2\alpha\hslash^2i}\left[J_0(r\kappaup^{1/\alpha}/\hslash)+iY_0(r\kappaup^{1/\alpha}/\hslash)\right]\equiv \frac{\kappaup^{(2-\alpha)/\alpha}}{2\alpha\hslash^2i}H^{(1)}_0(r\kappaup^{1/\alpha}/\hslash)
\\&\sim\frac{\kappaup^{(2-\alpha)/\alpha}}{2\alpha\hslash^2i}\sqrt{\frac{2\hslash}{\pi \kappaup^{1/\alpha}r}}\exp[(r\kappaup^{1/\alpha}/\hslash-\frac{\pi}{4})i] \\&=-\frac{\kappaup^{(3-2\alpha)/(2\alpha)}}{\alpha\sqrt{2\pi r\hslash^3}}e^{\frac{\pi}{4}i}\exp[ir\kappaup^{1/\alpha}/\hslash],\mbox{ when}
 r\rightarrow+\infty.\label{approxgreen}
\end{split}
\end{equation}
In a similar way to the time-dependent case,  using the first term of the above formula instead of the exact results for the Green's function $G^+(\mathrm{\textbf{r}})$,  the integral equation(\ref{green1}) can be simplified to
\begin{equation}
\phi(\mathrm{\textbf{r}})\cong
\mbox{e}^{i\mathrm{\textbf{k}}\cdot\mathrm{\textbf{r}}}-\frac{\kappaup^{(3-2\alpha)/(2\alpha)}e^{\frac{\pi}{4}i}}{\sqrt{2\pi r\hslash^3}\alpha D_\alpha}\int{e^{i\kappaup^{1/\alpha}|\mathrm{\textbf{r}}-\mathrm{\textbf{r}}_0|/\hslash}}
V(\mathrm{\textbf{r}}_0)\phi(\mathrm{\textbf{r}}_0)\mathrm{d}^2\mathrm{\textbf{r}}_0.\label{forborn1}
\end{equation} Meantime, we have
\begin{equation}
|\mathrm{\textbf{r}}-\mathrm{\textbf{r}}_0|=\sqrt{\mathrm{\textbf{r}}^2-2\mathrm{\textbf{r}}\cdot\mathrm{\textbf{r}}_0+\mathrm{\textbf{r}}_0^2}
\cong
r-\frac{\mathrm{\textbf{r}}\cdot\mathrm{\textbf{r}}_0}{r}.\label{approx}
\end{equation}Substituting
$|\mathrm{\textbf{r}}-\mathrm{\textbf{r}}_0|$ by Eq.~(\ref{approx})
in Eq.~(\ref{forborn1}) yields
\begin{equation}
\phi(\mathrm{\textbf{r}})\cong\phi_0(\mathrm{\textbf{r}})-\frac{\kappaup^{3/(2\alpha)-1}\mbox{e}^{i(\kappaup^{1/\alpha}r/\hslash+\pi/4)}}{\sqrt{2\pi r\hslash^3}\alpha D_\alpha}\int
\mbox{e}^{-i\mathrm{\textbf{k}}_f\cdot\mathrm{\textbf{r}}_0}V(\mathrm{\textbf{r}}_0)
\phi(\mathrm{\textbf{r}}_0)\mathrm{d}^2\mathrm{\textbf{r}}_0,\label{origin1}
\end{equation}
where
\begin{equation}\mathrm{\textbf{k}}_f=\frac{\kappaup^{1/\alpha}}{\hslash}\frac{\mathrm{\textbf{r}}}{r}.
\end{equation}  Now we invoke the Born approximation \cite{Shankar,born}:
Suppose the incoming plane wave is not substantially altered by the
potential. Then, in Eq.~(\ref{forborn1}), it makes sense to use
\begin{equation}
\phi(\mathrm{\textbf{r}}_0)\approx\phi_0(\mathrm{\textbf{r}}_0)=\mbox{e}^{i\mathrm{\textbf{k}}\cdot\mathrm{\textbf{r}}_0},\label{origin2}
\end{equation} which yields
\begin{equation}
\phi(\mathrm{\textbf{r}})\approx\mbox{e}^{i\mathrm{\textbf{k}}\cdot\mathrm{\textbf{r}}}-\frac{\kappaup^{3/(2\alpha)-1}\mbox{e}^{i(\kappaup^{1/\alpha}r/\hslash+\pi/4)}}{\sqrt{2\pi r\hslash^3}\alpha D_\alpha}\int
\mbox{e}^{-i\mathrm{\textbf{q}}\cdot\mathrm{\textbf{r}}_0}V(\mathrm{\textbf{r}}_0)\mathrm{d}^3\mathrm{\textbf{r}}_0,\label{born1}
\end{equation}
where
$\mathrm{\textbf{q}}=\mathrm{\textbf{k}}_f-\mathrm{\textbf{k}}$.
Here, vector $\mathrm{\textbf{k}}$ points in the incident direction,
$\mathrm{\textbf{k}}_f$ in the scattered direction, and they both
have magnitude $\kappaup^{1/\alpha}/\hslash$. Therefore,
$q=|\mathrm{\textbf{q}}|=2\kappaup^{1/\alpha}sin(\theta/2)/\hslash$,
where $\theta$ is the scattering angle, the angle between
$\mathrm{\textbf{k}}_f$ and $\mathrm{\textbf{k}}$. So,
$\hslash\mathrm{\textbf{q}}$ is the momentum transfer in the process
\cite{levin,Shankar}. Eq.~(\ref{born1}) gives the approximate wave
function of the fractional quantum scattering problem, and the
second term of the right side of this equation denotes the
scattering wave.
\par In the zeroth-order Born approximation the incident
plane wave passes by with no modification, and what we explored in
 Eq.~(\ref{born1}) can be viewed as the first-order correction to this.
We can also generate a series of higher-order corrections: the Born
series \cite{Shankar,born}. The generalized Lippmann-Schwinger
equation (\ref{lsequation1}) can be written as
\begin{equation}
\phi(\mathrm{\textbf{r}})=\phi_0(\mathrm{\textbf{r}})+\int
g(\mathrm{\textbf{r}}-\mathrm{\textbf{r}}_0)V(\mathrm{\textbf{r}}_0)\phi(\mathrm{\textbf{r}}_0)\mathrm{d}^3\mathrm{\textbf{r}}_0,\label{integral1}
\end{equation}
where $g(\mathrm{\textbf{r}})=G^+(\mathrm{\textbf{r}})/D_\alpha.$ We
can build an iteration scheme for the wave function as
\begin{equation}
\phi^{(n)}=\mbox{e}^{i\mathrm{\textbf{k}}\cdot\mathrm{\textbf{r}}}+\int
g(\mathrm{\textbf{r}}-\mathrm{\textbf{r}}_0)V(\mathrm{\textbf{r}}_0)\phi^{(n-1)}(\mathrm{\textbf{r}}_0)\mathrm{d}^3\mathrm{\textbf{r}}_0.\label{iteration}
\end{equation}Note that
$\phi^{(0)}=\mbox{e}^{i\mathrm{\textbf{k}}\cdot\mathrm{\textbf{r}}}$,
and $\phi^{(n)}$ is the $n$th-order corrections to the wave
function. For a given potential function, using
Eq.~(\ref{iteration}), the analytical approximate solutions of every
order can be obtained. In a series form, we have
\begin{equation}
\phi=\phi_0+\int gV\phi_0+\iint gVgV\phi_0+\iiint
gVgVgV\phi_0+\cdots.
\end{equation}In each integrand only the incident wave function ($\phi_0$)
appears, together with more and more powers of $gV$.
\section{ {Conclusions}}
In this paper, the 2D space-FSE with time-dependent and time-independent potentials was
studied. We defined the Green's function of the FSE for the fractional scattering problem in the two cases,
 and the FSE was converted into an integral form. We gave
the mathematical expression of the Green's function in terms of
some special functions. The asymptotic
properties of the Green's functions for
$|\mathrm{\textbf{r}}-\mathrm{\textbf{r}}^\prime|\rightarrow\infty$
(or $|\mathrm{\textbf{r}}|>>|\mathrm{\textbf{r}}^\prime|$) were also
given. Using these results, we obtained
the approximate scattering wave function for the fractional quantum scattering problems (see Eqs.~(\ref{TDapproxwave2}) and (\ref{born1})).
A series of higher-order corrections to the approximate wave
functions were also given in Eqs.~(\ref{TDiteration}) and (\ref{iteration}). These results are
useful for the time-dependent scattering problem in the fractional
quantum mechanics. All of these results contain those in the
standard quantum mechanics as special cases.

\section*{ {Acknowledgements}}
This work was supported by the National Natural Science Foundation of China (Grant
No. 11147109), the Specialized Research Fund for the Doctoral Program of Higher Education
of China (Grant No. 20113218120030), and the Fundamental Research Funds for the Central
Universities (Grant No. NS2012119).

\appendix
\renewcommand{\theequation}{A\arabic{equation}}
\setcounter{equation}{0}
\renewcommand{\thesection}{Appendix:}
\section{ {Fox's $H$-function and Some Properties}}
The Fox's $H$-function \cite{mathai0,mathai} is defined by an
integral of Mellin-Barnes type \cite{Paris} as
\begin{equation}H^{m,n}_{p,q}(z)=H^{m,n}_{p,q}\left[z\text{ }\Bigg|
\begin{aligned}&{\text{
}(a_1,A_1),(a_2,A_2),\ldots,(a_p,A_p)}
\\&{\text{
}(b_1,B_1),(b_2,B_2),\ldots,(b_p,B_p)}
\end{aligned}\right]=\cfrac1{2\pi
i}\int _{L}\chi(s)z^{-s}\mathrm{d}s,\label{a1}\end{equation}where
\begin{equation}
\chi(s)=\cfrac{\prod^m_{j=1}\Gamma(b_j+B_js)\prod^n_{i=1}\Gamma(1-a_i-A_is)}{\prod^p_{i=n+1}\Gamma(a_i+A_is)\prod^q_{j=m+1}\Gamma(1-b_j-B_js)}.\label{a1add}
\end{equation} The contour $L$ runs from $c-i\infty$ to $c+i\infty$ separating the
poles of $\Gamma(1-a_i-A_is)$, ($i=1,\cdots,n$) from those of
$\Gamma(b_j+B_js)$, ($j=1,\cdots,m$).  Here we present some
properties of the $H$-function used in our paper. In order to give
the results, the following definitions will be used,
\begin{equation}
\begin{aligned}
&\Delta=\sum^q_{j=1}B_j-\sum^p_{i=1}A_j;\quad\Delta^*=\sum^n_{i=1}A_i-\sum^p_{i=n+1}A_i+\sum^m_{j=1}B_j-\sum^q_{j=m+1}B_j;
\\&\delta=\prod^p_{j=1}(A_j)^{-A_j}\prod^q_{j=1}(B_j)^{B_j};
\quad\mu=\sum^q_{j=1}b_j-\sum^p_{i=1}a_i+\frac{p-q}{2}.
\end{aligned}\label{a2}
\end{equation}The following properties of the  $H$-function can be
found in Refs.~\cite{mathai,mathai0,kilbas0}.
\\\textbf{Property 1:}\\
\begin{align}
\frac{1}{k} H^{m,n}_{p,q}\left[z\text{ }\Bigg|
\begin{aligned}&{\text{
}(a_p,A_p)}
\\&{\text{
}(b_p,B_p))}
\end{aligned}\right]=H^{m,n}_{p,q}\left[z^k\text{ }\Bigg|
\begin{aligned}&{\text{
}(a_p,kA_p)}
\\&{\text{
}(b_p,kB_p)}
\end{aligned}\right].\quad\mbox{for } k>0\label{hproperty1}
\end{align}
\textbf{Property 2:}\\
\begin{align}
z^\sigma H^{m,n}_{p,q}\left[z\text{ }\Bigg|
\begin{aligned}&{\text{
}(a_p,A_p)}
\\&{\text{
}(b_p,B_p))}
\end{aligned}\right]=H^{m,n}_{p,q}\left[z\text{ }\Bigg|
\begin{aligned}&{\text{
}(a_p+\sigma A_p,A_p)}
\\&{\text{
}(b_p+\sigma B_p,B_p)}
\end{aligned}\right].\quad\mbox{for } \sigma\in\mathbb{C}\label{hproperty2}
\end{align}
\textbf{Property 3: Explicit Power Series Expansion }\par For
$\Delta>0$ , $z\neq0$ or $\Delta=0$, $|z|>\delta$, there holds the
following expansion for the $H$-function \cite{mathai},
\begin{align}
H^{m,n}_{p,q}(z)=\sum^{m}_{h=1}\sum^{\infty}_{k=0}\cfrac{\prod^m_{j=1,j\neq
h}\Gamma(b_j-B_js_{hk})\prod^n_{i=1}\Gamma(1-a_i+A_is_{hk})}
{\prod^p_{i=n+1}\Gamma(a_i-A_is_{hk})\prod^q_{j=m+1}\Gamma(1-b_j+B_js_{hk})}\frac{(-1)^k}{k!}\frac{z^{s_{hk}}}{B_h},\label{hproperty3}
\end{align} where $s_{hk}=(b_h+k)/B_h$,
if the following conditions are satisfied:
\begin{enumerate}
\item The poles of the gamma functions $\Gamma(1-a_i-A_is)$, ($i=1,\cdots,n$) and those
of $\Gamma(b_j+B_js)$, ($j=1,\cdots,m$) do not coincide:
\begin{equation}
A_i(b_j+l)\neq B_j(a_i-k-1), (i=1,\cdots,n;
j=1,\cdots,m;k,l=0,1,2,\cdots).\label{condition1}
\end{equation}
\item The poles of the gamma functions $\Gamma(b_j+B_js)$,
($j=1,\cdots,m$) are simple:
\begin{equation}
B_i(b_j+l)\neq B_j(b_i+k), (i\neq
j;i,j=1,\cdots,m;k,l=0,1,2,\cdots).\label{condition2}
\end{equation}
\end{enumerate}
\ \textbf{Property 4: Asymptotic Expansions at Infinity in the Case
$\Delta>0$, $\Delta^*=0$ }\par When the poles of the gamma functions
$\Gamma(1-a_i-A_is)$, ($i=1,\cdots,n$) are simple:
\begin{equation} A_i(1-A_j+l)\neq
A_j(1-a_i+k), (i\neq
j;i,j=1,\cdots,n;k,l=0,1,2,\cdots),\label{condition3}
\end{equation} and the condition (\ref{condition1})
is satisfied, the $H$-function has the following asymptotic
expansion \cite{kilbas,kilbas0},
\begin{equation}
\begin{split}
H^{m,n}_{p,q}(z)=&\sum^{n}_{i=1}\left[h_iz^{(a_i-1)/A_i}+\mbox{o}\left(z^{(a_i-1)/A_i}\right)\right]+Az^{(\mu+1/2)/\Delta}\Big(c_0\exp\left[(B+Cz^{1/\Delta})i\right]-
\\&d_0\exp\left[-(B+Cz^{1/\Delta})i\right]\Big)+\mbox{o}\left(z^{(\mu+1/2)/\Delta}\right),\label{hproperty4}
\end{split}
\end{equation}
where \begin{align}
&h_i=\frac{1}{A_i}\cfrac{\prod^m_{j=1}\Gamma(b_j-B_j(a_i-1)/A_i)\prod^n_{j=1,j\neq
i}\Gamma(1-a_j+A_j(a_i-1)/A_i)}
{\prod^p_{j=n+1}\Gamma(a_j-A_j(a_i-1)/A_i)\prod^q_{j=m+1}\Gamma(1-b_j+B_j(a_i-1)/A_i)},\label{hp4hi}
\\&A=\frac{A_0}{2\pi
i\Delta}\left(\frac{\Delta^\Delta}{\delta}\right)^{(\mu+1/2)/\Delta},\quad
B=\frac{(2\mu+1)\pi}{4}, \quad
C=\left(\frac{\Delta^\Delta}{\delta}\right)^{1/\Delta},\label{hp4abc}
\\&A_0=(2\pi)^{(p-q+1)/2}\Delta^{-\mu}\prod^p_{i=1}A_i^{-a_i+1/2}\prod^q_{j=1}B_j^{b_j-1/2},\label{hp4a0}
\\&c_0=(2\pi
i)^{m+n-p}\exp\left[\left(\sum^p_{i=n+1}a_i-\sum^m_{j=1}b_j\right)\pi
i\right],\label{hp4c0}
\\\mbox{and } \quad &d_0=(-2\pi
i)^{m+n-p}\exp\left[-\left(\sum^p_{i=n+1}a_i-\sum^m_{j=1}b_j\right)\pi
i\right].\label{hp4d0} \end{align}

\end{document}